\documentclass[showpacs,amsmath,amssymb,superscriptaddress]{revtex4}

\usepackage{graphicx}
\usepackage{bm}
\usepackage{color}

\begin{document}

\title{Residual mean first-passage time for jump processes: theory
and applications to L{\'e}vy flights and fractional Brownian motion}

\author{V. Tejedor}
\affiliation{Laboratoire de Physique Th\'eorique de la Mati\`ere Condens\'ee
(UMR 7600), Universit\'e Pierre et Marie Curie, 4 Place Jussieu, 75255
Paris Cedex}
\affiliation{Physics Department, Technical University of Munich, James Franck
Strasse, 85747 Garching, Germany}
\author{O. B\'enichou}
\affiliation{Laboratoire de Physique Th\'eorique de la Mati\`ere Condens\'ee
(UMR 7600), Universit\'e Pierre et Marie Curie, 4 Place Jussieu, 75255
Paris Cedex}
\author{Ralf Metzler}
\affiliation{Physics Department, Technical University of Munich, James Franck
Strasse, 85747 Garching, Germany}
\affiliation{Physics Department, Tampere University of Technology,
FI-33101 Tampere, Finland}
\author{R. Voituriez}
\affiliation{Laboratoire de Physique Th\'eorique de la Mati\`ere Condens\'ee
(UMR 7600), Universit\'e Pierre et Marie Curie, 4 Place Jussieu, 75255
Paris Cedex}

\date{\today}

\begin{abstract}
We derive a functional equation for the mean first-passage time (MFPT) of a
generic self-similar Markovian continuous process to a target in a
one-dimensional domain and obtain its exact solution. We show that the obtained
expression of the MFPT for continuous processes is actually different from the
large system size limit of the MFPT for discrete jump processes allowing
leapovers. In the case considered here, the asymptotic MFPT admits non-vanishing
corrections, which we call \emph{residual MFPT}. The case  of L{\'e}vy flights
with diverging variance of jump lengths is investigated in detail, in particular,
with respect to the associated leapover behaviour. We also show numerically that
our results apply with good accuracy to fractional Brownian motion, despite its
non-Markovian nature.
\end{abstract}


\maketitle

\section{Introduction}


The  first-passage time (FPT) of a random walker to a target point is an important concept for the modelling and interpretation of
stochastic processes \cite{Hughes1995,Redner:2001}. This interest in the concept of FPTs
is motivated by the crucial role played by FPTs in various contexts, including
diffusion limited chemical reactions \cite{Rice:1985,Hanggi:1990a,PCCP}, the spreading of diseases
\cite{Lloyd:2001}, or target search processes
\cite{Benichou:2005qd,Benichou:2006lq,Shlesinger:2006,Lomholt:2008}. 
A general formalism for the calculation of the mean first-passage time (MFPT) of
a scale-invariant random process ${\bf r}_t$ in a confined domain of volume $V$
has recently been derived in the large $V$ limit in \cite{Condamin:2007zl,Tejedor:2009vn}, and
the full distribution of the FPT has been obtained \cite{BenichouO.:2010a}.
These results highlight the asymptotic dependence of the MFPT ${\langle{\bf T}
\rangle}$ and its  higher order moments on both the source-to-target distance
$r$ and on the confinement volume in the case of scale-invariant processes,
which can be characterized by the walk dimension $d_w$ (defined by $\langle
{\bf r}_t^2\rangle\propto t^{2/d_w}$) and the fractal dimension $d_f$ of the
support of the random process \cite{havlin}. In particular, in the case of compact (i.e.,
recurrent) processes, for which $d_w>d_f$, it was shown that
\begin{equation}\label{scale}
\lim_{V\to\infty}\frac{\langle{\bf T}\rangle}{V}=Br^{d_w-d_f},
\end{equation}
where only the constant $B$ depends on the process.
For a Brownian walker the first passage at a given point $x$, which hereafter denotes 
the first \emph{arrival at exactly\/}  $x$ is
equivalent to the first \emph{crossing\/} of the value $x$. This is not
generally true. For instance, in the case of L{{\'e}vy flights with
scale-free distribution of jump lengths \cite{bouchaud,mekla,R.Metzler:2004}, pronounced leapovers
across the threshold value occur. Thus, the first passage at the point $x$
becomes considerably less likely than the first crossing of the value $x$ \cite{Chechkin:2003}.
For Markovian L{\'e}vy flights, the first crossing falls into the class of
processes governed by the Sparre Andersen universality \cite{Metzler:2007}.

Discrete time Markovian jump processes are defined by an elementary jump
distribution $w({\bf r})$, which, at each discrete time step $n$, renders the
probability that the random walker makes a jump of length ${\bf r}$, which we
here assume to be discrete. For a generic distribution $w$, the searcher is in
principle allowed to jump across the target, and perform what we define here as a leapover, as opposed to the case of nearest
neighbor random walks. In the case of Euclidean spaces, symmetric jump processes
are subject to the generalized central limit theorem
\cite{Polya:1920,Kolmogorov:1954}, and have a well defined continuous limit
which is scale-invariant. More explicitly, let $P({\bf r},n)$ denote the
discrete time and space propagator of the process in infinite space, starting
from ${\bf r}=0$ at time $n=0$. Then we know that for the rescaled quantity
\begin{equation}
p_\epsilon({\bf x} ={\bf r}\epsilon,t=n\epsilon^{d_w})\equiv\frac{1}{\epsilon}
P({\bf r},n),
\end{equation}
 the convergence to
\begin{equation}\label{contprocess}
\lim_{\epsilon\to0}p_\epsilon({\bf x} ={\bf r}\epsilon,t =n\epsilon^{d_w})=
p({\bf x},t)\equiv t^{-d_f/d_w}g({\bf x}/t^{1/d_w})
\end{equation}
is fulfilled in the continuum limit $\epsilon\to0$
with ${\bf x}$ and $t$ fixed by virtue of the central limit theorem. Here $g$ is a scaling function. If the jump distribution $w({\bf
r})$ has a second moment $\langle\mathbf{r}^2\rangle$, the limit process is a
Brownian motion with $d_w=2$ and $g$ is Gaussian; otherwise the limit process
is of L{\'e}vy stable type with $0<d_w<2$, and $g$ is a L{\'e}vy stable law of
index $d_w$. In this continuous limit the asymptotic theory applies
\cite{Condamin:2007zl}, for compact processes yielding the exact asymptotic
form (\ref{scale}), where $B$ can be explicitly determined as a function of
$g$. These results show in particular that the MFPT in the continuous limit
becomes {\it independent} of the existence of leapovers.

In terms of the variables ${\bf r}$ and $n$ of the discrete jump process, the
continuum limit is equivalent to taking both ${\bf r}$ and $n$ large, with
${\bf r}^{d_w}/n$ fixed. Hence, for discrete jump processes, Eq.~(\ref{scale})
produces actually only the leading term of a large $r$ expansion, and one
should write more precisely
\begin{equation}\label{scalebis}
\lim_{V\to\infty}\frac{\langle {\bf T} \rangle}{V}=Br^{d_w-d_f} +o(r^{d_w-d_f}).
\end{equation} 
We show in this paper that the subleading term of this expansion actually
strongly depends on the existence of leapovers, and may yield non-vanishing
corrections to the rescaled MFPT. This \emph{residual MFPT\/} can have
important consequences in the context of both search processes and numerical
simulations of first-passage times of random walks.

As a specific important example we discuss extensively the effect of leapovers
on the MFPT in the context of one-dimensional ($1D$) Markov jump processes. By
means of a functional equation, we first derive an exact expression of the MFPT
for continuous time and space scale-invariant Markov processes valid for any
starting position of the walker and confining domain size. Next, using the
method of pseudo Green functions, we present an exact expression of the MFPT
for arbitrary discrete time $1D$ Markov jump processes, which was previously
derived in Ref.~\cite{Hughes1995,Montroll:1965}. We show that in the large
volume limit, the leading term of the MFPT is the expected continuous limit of
Eq.~(\ref{scale}), and we calculate exactly the subleading term of this
expansion. We show that this correction is indeed crucial in the case of random
walks with leapovers, since it does not vanish in the limit $L\to\infty$ and
$r\to 0$.

We consider a walker performing random independent jumps in a confined $1D$
system of size $L$ (with periodic boundary conditions), and address the
question of determining its mean first-passage time $\langle {\bf T}_{TS}
\rangle\equiv\langle {\bf T}(r,L)\rangle$ to a given target site $T$ as a
function of its starting site $S$, where $r$ stands for the source-to-target
distance $|ST|$. 
To proceed, we denote by
$w_{ji}$ the jump probability from discrete site $i$ to $j$, where $w_{ji}$ is assumed
to be symmetric ($w_{ji}=w_{ij}$) and translation-invariant ($w_{ji}$ is a function $|i-j|$ only).
It is easily seen that the confined problem with periodic boundary conditions
is equivalent to an infinite line with regularly spaced targets at positions
$k L$ with $k\in\mathbb{Z}$, a property which will be used below.

\section{Continuous time and space  scale-invariant jump processes} 

In this section we derive an exact expression for the MFPT of a $1D$
scale-invariant Markov jump processe in the limit of continuous space and
time. We start from the definition in discrete time and space given above,
and pass to the following classical backward equation for the MFPT
\cite{Redner:2001}:
\begin{equation}\label{back}
\sum_{i=-\infty}^{\infty}\langle {\bf T}_{Ti} \rangle(w_{iS}-\delta_{iS})\equiv
\Delta_S \langle {\bf T}_{TS} \rangle=-1, 
\end{equation}
completed by the condition $\langle {\bf T}(r=kL,L) \rangle=0$ for all $k \in
\mathbb{Z}$.
It is easily seen that the following form of the MFPT
\begin{equation}
\langle {\bf T} (r,L) \rangle = \frac{\langle {\bf T} (r,2L) \rangle +
\langle {\bf T} (r+L,2L) \rangle - \langle {\bf T} (L,2L) \rangle}{2}
\end{equation}
satisfies both Eq.~(\ref{back}) and the boundary conditions above. Note that by
definition, $\langle {\bf T} (r,L) \rangle$ is $L$-periodic. This provides a
functional equation, which can be solved as follows. We assume that the jump
process has a well defined, self-similar continuous limit, characterized by a
walk dimension $d_w$. We then define the continuous limit of the MFPT as
\begin{equation}\label{continuous}
\theta(y,l)=\lim_{\epsilon\to 0} \epsilon^{d_w}\langle {\bf T}(r=y/\epsilon,
L=l/\epsilon) \rangle,
\end{equation}
where the starting position $y$ and system size $l$ are fixed. In particular,
we assume that the process is compact ($d_w > d_f = 1$), such that in the
continuous limit the MFPT to a point-like target is finite. In this limit
self-similarity implies
\begin{equation}
\theta(y,l) = l^{d_w}\theta(x=y/l,1),
\end{equation}
and therefore
\begin{equation}
\theta (x,1) = 2^{d_w-1} \left ( \theta (\frac{x}{2},1)  + \theta
(\frac{x+1}{2},1) - \theta (\frac{1}{2},1) \right ),
\end{equation}
so that we now have to solve the following functional equation
\begin{equation}
f (x) = 2^{d_w-1} \left ( f \left (\frac{x}{2} \right ) + f \left
(\frac{x+1}{2} \right ) - f \left (\frac{1}{2} \right ) \right ).
\label{eq:f}
\end{equation}
We know that $f(x)=f(1-x)$ by symmetry, and that $f(0)=f(1)=0$. Differentiation
of this equation yields
\begin{equation}
f'(x) = 2^{d_w-2} \left ( f '\left (\frac{x}{2} \right ) +
f' \left (\frac{x+1}{2} \right )\right ),
\label{eq:fp}
\end{equation}
together with the symmetry condition $f'(x)=-f'(1-x)$. It can be checked
directly that equation (\ref{eq:fp}) is solved by the Hurwitz zeta function
$\zeta(2-d_w,x) $, the latter being defined by the series
\begin{equation}
\displaystyle \zeta (s,q) = \sum_{n=0}^{\infty} \frac{1}{(q+n)^s}.
\end{equation}
Note that $\zeta(2-d_w,1- x)$ is also a solution. In order to satisfy the
symmetry condition, a solution of Eq.~(\ref{eq:fp}) satisfying $f'(x)=-f'(1-x)$
is $\zeta(2-d_w,x)-\zeta(2-d_w,1- x)$. A solution of the original equation
(\ref{eq:f}) satisfying $f(0)=0$ is then
\begin{equation}
f(x) = \int_0^x \left(\zeta (2-d_w,y) -  \zeta (2-d_w,1- y) \right)dy.
\label{eq:f2}
\end{equation}
Furthermore, using the fact that the Hurwitz zeta function satisfies the
relation
\begin{equation}
\frac{\partial \zeta (s,x)}{\partial x} =  -s \zeta (s+1,x),
\end{equation}
one can simplify Eq.~(\ref{eq:f2}) and write
\begin{equation}
f(x)=\frac{1}{d_w-1}\left ( \zeta (1-d_w,x) + \zeta (1-d_w,1-x) -
\zeta (1-d_w,0) -\zeta (1-d_w,1)\right ).
\label{eq:f3}
\end{equation}
The MFPT eventually becomes
\begin{equation}
\theta(y,l)  = A L^{d_w} \left ( \zeta (1-d_w,\frac{y}{l}) + \zeta
(1-d_w,1-\frac{y}{l}) - \zeta (1-d_w,0) -\zeta (1-d_w,1) \right),
\label{eq:f4}
\end{equation}
where $A$ is a multiplicative constant which remains undetermined at this stage.
For $d_w=2$ (Brownian case), we retrieve the classical expression, using $\zeta
(-n,x) = -B_{n+1}(x)/(n+1)$, where $B_n$ are the Bernoulli polynomials ($B_2(x)
= x^2-x+1/6$):
\begin{eqnarray*}
f(x)  & = & \zeta (-1,x) + \zeta (-1,1-x)  - \zeta (-1,0) -\zeta (-1,1)\\
& = & -\frac{1}{2} \left ( B_2(x) +B_2(1-x)  -B_2(0) -B_2(1) \right )\\
& = & x (1-x).
\end{eqnarray*}
Comparison of this expression to the well known result $\theta(y,l)=y(l-y)$
yields $A=1$ in this Brownian case, with $D=1/2$.

In the general case $d_w\not=2$, $A$ can be calculated by using the exact
asymptotic result in the large $l$ limit of Eq.~(\ref{scale}) (see
Ref.~\cite{Condamin:2007zl} for details). For $y\ll l$, the asymptotic behavior of
Eq.~(\ref{eq:f4}) readily produces
\begin{equation}\label{asympcontinuous}
\lim_{l\to\infty} \frac{\theta(y,l)}{l}=A  y^{d_w-1}.
\end{equation}
Comparison with the general result of Ref.~\cite{Condamin:2007zl} then
delivers
\begin{equation}
A = \int_0^{\infty} \frac{du}{u^{d_f/d_w}} g^* \left ( u^{-1/d_w} \right ),
\label{eq:defA}
\end{equation}
where the infinite space propagator of the process $p(r,t)$ satisfies:
\begin{equation}
p(r,t) = t^{-1/d_w} g( r/t^{1/d_w}),
\end{equation}
and $g^*(u) = g(0) - g(u)$ tends to $0$ when $u \to 0$.

Because of the generalized central limit theorem invoked in the introduction,
a generic continuous time self-similar Markov process with $d_w\neq2$ is a
L{\'e}vy process, whose infinite space propagator reads
\begin{equation}\label{levyprop}
p(r,t) = \frac{1}{\pi} \int_0^{\infty} \cos (r x) e^{\displaystyle -
t (c x )^{d_w}} dx,
\end{equation}
where $c$ is a scaling parameter. We subsequently deduce
\begin{equation}
g (u) = \frac{1}{\pi} \int_0^{\infty} \frac{1}{u} \cos (v) e^{\displaystyle -
(\frac{c v}{u} )^{d_w}} dv,
\end{equation}
and
\begin{equation}
g^*(u) = \frac{1}{\pi} \int_0^{\infty} \frac{1}{u} \left ( 1 - \cos (v) \right ) e^{\displaystyle -  (\frac{c v}{u} )^{d_w}} dv.
\end{equation}
The constant $A$ can then be explicitly calculated and reads:
\begin{eqnarray}
\label{Alevy}
A & = & \int_0^{\infty} \frac{du}{\pi u^{1/d_w}} \int_0^{\infty} u^{1/d_w}
\left ( 1 - \cos (v) \right ) e^{\displaystyle -  (c v )^{d_w} u} dv
\nonumber \\
& = & \int_0^{\infty} \frac{1 - \cos (v)}{\pi (c v )^{d_w}} dv \nonumber \\
A & = & \frac{d_w }{\displaystyle  2 c^{d_w} \Gamma(1+d_w) \cos \left (
\frac{(2-d_w) \pi}{2} \right )}. \label{eq:A}
\end{eqnarray}
Equations (\ref{eq:f4}) and (\ref{Alevy}) provide an explicit and exact
expression of the MFPT for all $r$ and $L$, valid for any continuous time
and space scale-invariant Markov process.

In order to check equation (\ref{eq:f4}), we simulated L{\'e}vy flights with
various values of $d_w$. A L{\'e}vy flight is here modeled as a discrete-time
random walk, where each step-size is a random integer $x$, distributed according
to the distribution
\begin{equation}
w(x = n) = \frac{1}{2 \zeta(1+d_w) |n|^{1+d_w}},
\label{eq:distrib}
\end{equation}
with $w(x=0)=0$. The random walk takes place on a $1D$ lattice of size $N$,
in discrete time, and with periodic boundary conditions. The target is located
at the site of origin $0\equiv N$. This processe converges in the continuous
limit defined in introduction to a L{\'e}vy process, whose infinite space
propagator is given by Eq.~(\ref{levyprop}) with
\begin{equation}
c^{d_w} = \frac{\pi}{\displaystyle 2 d_w \sin \left ( \frac{\pi d_w}{2} \right
)\Gamma (d_w) \zeta(1 + d_w)}. \label{eq:c}
\end{equation}

We compared the exact expression (\ref{eq:f4}) obtained in the continuous limit
with our discrete simulation. Figs.~\ref{fig:LevyT.25} and \ref{fig:LevyT.5}
show that as the lattice size $N$ grows, the simulated discrete MFPT approaches
the exact continuous limit, for different values of $d_w$ (for each $d_w$, the
multiplicative constant $A$ is computed using (\ref{eq:A})). Note the very slow
convergence to the exact solution in the continuous limit. We show below that
leapovers lead to non vanishing corrections of the MFPT to this continuous
limit in the large system size limit, which we define as the residual MFPT.

 \section{Discrete time and space  jump processes}
 
\begin{figure}
\includegraphics[width =0.5\linewidth,clip]{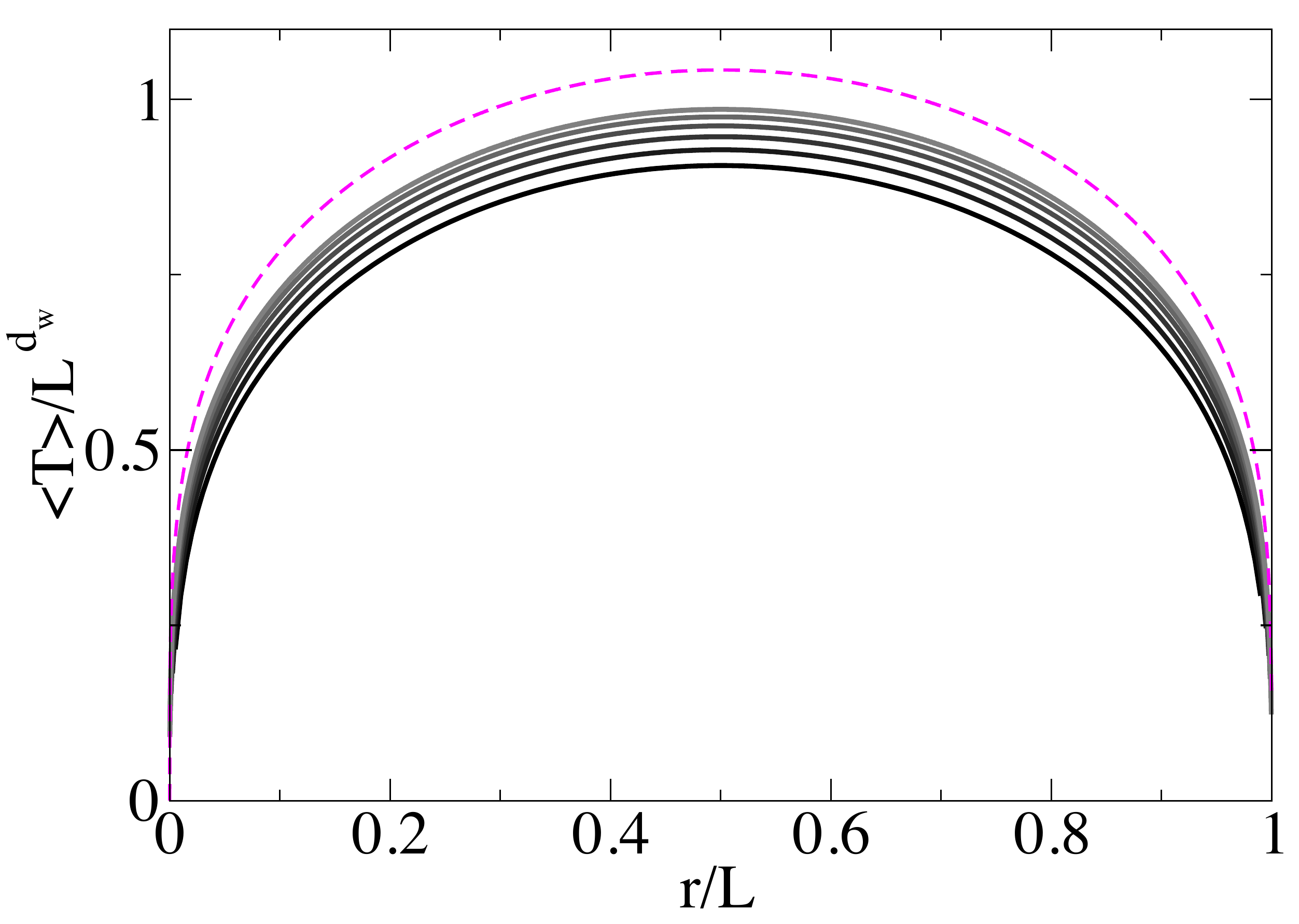} 
\caption{Mean First Passage Time for a L{\'e}vy flight, for several sizes of the $1D$ lattice (200, 400, 800, 1600, 3200 and 6400 sites, from black [bottom] to light grey [top] as the size grows) compared to the theoretical expression
(\ref{eq:f4}) (magenta dotted line, the $A$ constant being given by
Eq.~(\ref{eq:A})), and for $d_w = 1.25$.}
\label{fig:LevyT.25}
\end{figure}

\begin{figure}
\includegraphics[width =0.5\linewidth,clip]{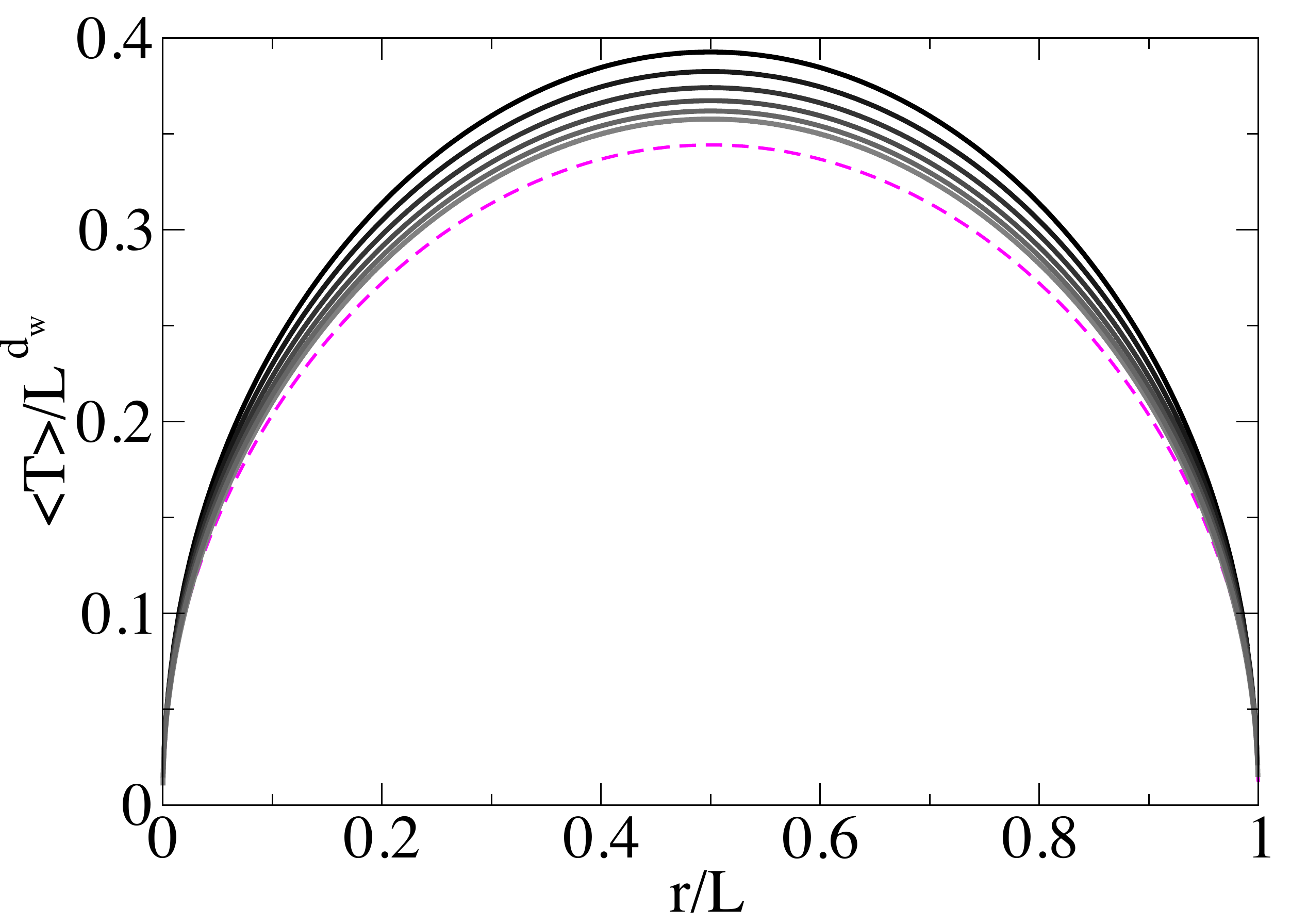} 
\caption{Mean First Passage Time for a L{\'e}vy flight, for several size of the $1D$ lattice (200, 400, 800, 1600, 3200 and 6400 sites, from black [top] to light grey [bottom] as the size grows) compared to the theoretical expression of Eq. (\ref{eq:f4}) (magenta dotted line, the $A$ constant being given by Eq. (\ref{eq:A})), and for $d_w = 1.5$.}
\label{fig:LevyT.5}
\end{figure}


To investigate the convergence of the discrete process to its continuous limit,
we compute analytically the MFPT for a generic discrete time and space Markov
jump process, regarding jump processes with and without second moments. We use
the formalism developed by Montroll \cite{Montroll:1965} (see also Hughes
\cite{Hughes1995}). First we define the structure function $\lambda (k)$, the
Fourier transform of the jump  distribution (\ref{eq:distrib}):
\begin{equation}
\lambda(k)=\sum_{n=-\infty}^{\infty}w(x=n)e^{ink}.
\end{equation}
Denoting by $\tilde{P}_t(k)$ the Fourier transform of the propagator at discrete
time $t$ in infinite space, we obtain $\tilde{P}_t(k)=\lambda(k)^t$. The
infinite space Green's function then reads
\begin{eqnarray}
G_{ji} & = & \sum_{t=0}^{\infty} P(j,t|i,0) =  \sum_{t=0}^{\infty} \frac{1}{\pi} \int_{0}^{\pi} \cos (k |i-j|)  \lambda(k)^t dk \nonumber
\\& = & \frac{1}{\pi} \int_{0}^{\pi} \frac{\cos (k |i-j|)}{1- \lambda(k)} dk.
\end{eqnarray}
Recalling that there exists a set $\Sigma$ of targets located at sites $n L$
($n\in\mathbb{Z}$) it is useful to define
\begin{eqnarray}
G_{\Sigma r}&=&\sum_{n=-\infty}^{\infty}\frac{1}{\pi}\int_{0}^{\pi}\frac{\cos
(k (r+nL))}{1- \lambda(k)} dk \\
&=&\frac{1}{L} \sum_{m=1}^{L-1} \frac{\cos \left ( \frac{2
\pi m r}{L} \right )}{1- \lambda \left ( \frac{2 \pi m}{L} \right )}.
\label{eq:G}
\end{eqnarray}
It can then be checked directly that the following expression
\begin{equation}
\langle {\bf T} (r,L) \rangle = L(G_{\Sigma\Sigma}-G_{\Sigma r})= 
\sum_{m=1}^{L-1} \frac{1 - \cos \left ( \frac{2 \pi m r}{L}
\right )}{1- \lambda \left ( \frac{2 \pi m}{L} \right )}
\label{eq:Texact}
\end{equation}
represents an exact solution for the MFPT of a $1D$ discrete time and discrete
space symmetric random walk, since it satisfies both Eq.~(\ref{back}) and the
boundary condition  $\langle {\bf T}(r=nL,L)\rangle=0$ for all $n\in\mathbb{Z}$.
In what follows we distinguish (i) the continuous limit $\theta(y,l)$ of this
exact expression, defined by  Eq.(\ref{continuous}), and (ii) the scaled MFPT
in the large $L$ limit, which reads according to Eq.~(\ref{eq:Texact})
\begin{equation}
\lim_{L\to \infty}\frac{\langle {\bf T} (r,L) \rangle }{L}\equiv\tau(r)=
\frac{1}{\pi} \int_{0}^{\pi} \frac{1 - \cos \left ( k r \right )}{1-
\lambda \left (k \right )} dk.
\label{eq:Tapprox}
\end{equation}
While it is clear that the continuous limit $\theta(y,l)$ must yield the exact
result of Eq.~(\ref{eq:f4}) derived independently above, we will show below
that the scaled MFPT $\tau(r)$ can differ from this continuous limit as soon
as the jump distribution allows leapovers. The residual MFPT corresponds to the
non-vanishing, subleading term of $\tau(r)$ in the large $r$ regime, which can
be obtained by considering the $k \to 0$ behavior of $\lambda(k)$. We consider
different examples below, corresponding to jump distributions with and without
second moment.

\subsection{Residual MFPT for jump distributions with second moment.}

We first consider a random walk whose jump distribution possesses a second
moment. As a concrete example we employ the exponential law defined by
\begin{equation}
P(n) = \frac{(e^{\beta}-1)}{2} e^{- \beta |n|}
\end{equation}
for $n\ge1$ and $P(0)=0$, where $\beta$ is a scaling parameter. The second moment of this jump
distribution becomes
\begin{equation}
\langle n^2 \rangle=\sum_{n=-\infty}^{\infty} n^2 P(n)=\frac{e^{\beta}
(1+e^{\beta})}{(e^{\beta}-1)^2},
\end{equation}
so that the continuous limit of this jump process, according to
Eq.~(\ref{contprocess}), is a Brownian motion with diffusion coefficient
$D=\langle n^2 \rangle/2$, and therefore $d_w=2$. In this continuous limit
one obtains straightforwardly the following form of the MFPT,
\begin{equation}\label{expcont}
\theta(y,l)=\lim_{\epsilon\to 0} \epsilon^{d_w}\langle {\bf T} (r=y/\epsilon,
L=l/\epsilon) \rangle=\frac{(e^{\beta}-1)^2}{e^{\beta} (1+e^{\beta})} y (l-y).
\end{equation}
We now  apply the formalism of the previous section and first compute the
structure function $\lambda(k)$, obtaining
\begin{equation}
\lambda(k)=\sum_{n=-\infty}^{\infty}P(x=n) e^{ink}=(e^{\beta}-1)\sum_{n=1}^{
\infty} \cos ( nk) e^{-\beta n}= (e^{\beta}-1) \frac{e^{\beta} \cos(k)-1}{
\left( e^{\beta} \cos(k)-1 \right )^2 + \left( e^{\beta} \sin(k) \right )^2}.
\end{equation}
An exact expression of the MFPT is then given by (\ref{eq:Texact}), for which
we find
\begin{eqnarray}
\langle {\bf T}(r,L) \rangle & = &  \sum_{m=1}^{L-1} \frac{1
- \cos \left ( \frac{2 \pi m r}{L} \right )}{1- \lambda \left ( \frac{2 \pi
m}{L} \right )}\nonumber \\
& = & \frac{1}{e^{\beta} (e^{\beta}+1)} \sum_{m=1}^{L-1}
\frac{1 -\cos \left ( \frac{2 \pi m r}{L} \right )}{1- \cos \left ( \frac{2
\pi m}{L} \right )}\nonumber\\
& & \times \left ( (e^{\beta} - 1)^2 + 2 e^{\beta}(1- \cos \textstyle \left(
\frac{2 \pi m}{L} \right )) \right ) \nonumber\\
& = &  \frac{(e^{\beta} - 1)^2}{e^{\beta} (e^{\beta}+1)} r (L-r)\nonumber \\
& & + \frac{2 e^{\beta}}{e^{\beta} (e^{\beta}+1)} L .
\label{exactexp}
\end{eqnarray}
While this result has the expected continuous limit (\ref{expcont}), it reveals
a significant difference from the scaled MFPT, containing a non-vanishing
residual, subleading correction. More precisely we find that
\begin{equation}
\lim_{L\to \infty}\frac{\langle {\bf T}(r,L) \rangle}{L}= \tau(r)=\frac{(e^{
\beta} - 1)^2}{e^{\beta} (e^{\beta}+1)} r + \frac{2 e^{\beta}}{e^{\beta} (e^{
\beta}+1)} , 
\end{equation}
where we define the non-vanishing subleading term $2 e^{\beta}/(e^{\beta}
(e^{\beta}+1))$ as the residual MFPT. 

Actually, the residual MFPT for a generic jump distribution with second moment
can be fully derived. We write the small $k$ expansion of the structure function
as $\lambda(k)\sim 1-Dk^2$, where the diffusion coefficient is given by $D=
\langle n^2 \rangle/2$ as above. Then one has
\begin{eqnarray}
\tau(r)&=&\frac{1}{\pi} \int_{0}^{\pi} \frac{1 - \cos \left ( k r \right )}{
D k^{2}} dk  + B+o(1)\nonumber\\
&=&\frac{r}{2D} + B+o(1),
\end{eqnarray}
where the residual MFPT is given by
\begin{equation}\label{residualB0}
B=\frac{1}{\pi}\int_0^{\pi} \left(\frac{1}{1- \lambda \left (k \right )}-
\frac{1}{D k^{2}} \right)  dk -  \frac{1}{D \pi^{2}} .
\end{equation}
This reveals the crucial difference between the large system size limit on the
one hand and the continuous space and time limit for MFPTs of jump processes on
the other. Our above analysis (see Eq. (\ref{residualB0})) suggests that a
residual MFPT exists in the large system size limit as soon as leapovers are
allowed. In particular the residual MFPT depends on the full jump distribution
and not only on its second moment. For example, the case of a nearest neighbor
random walk, for which  $\lambda(k)=\cos(k)$, yields a vanishing residual MFPT,
as can be checked from Eq. (\ref{residualB0}).

\subsection{Residual MFPT for L{\'e}vy flights}

We now turn to jump distributions with infinite second moment. The continuous limit
of such jump processes is a L{\'e}vy flight by virtue of the generalized central
limit theorem. We start with the example of discrete space L{\'e}vy flights,
defined by the distribution (\ref{eq:distrib}), for which one has
\begin{equation}
\lambda(k)=\sum_{n=-\infty}^{\infty}w(x=n) e^{i n k}= \frac{1}{\zeta(d_w+1)}
\sum_{n=1}^{\infty} \frac{\cos ( nk)}{n^{d_w+1}}.
\end{equation}
We first note that using this expression in the exact result (\ref{eq:Texact})
yields in the continuous limit the result (\ref{eq:f4}) and (\ref{Alevy}), as
expected. We consider below the large $L$ limit of the MFPT and will therefore
make use of the following small $k$ expansion \cite{Hughes1995}
\begin{equation}
\lambda (k)  = 1 -  \alpha_1 k^{d_w}  + \alpha_2 k^2 + \mathcal{O} (k^{4}),
\end{equation}
where
\begin{equation}
\alpha_1 = \frac{\pi}{2 \zeta (d_w +1) \Gamma ( d_w + 1) \sin \left (
\frac{\pi d_w}{2} \right )},
\end{equation}
and
\begin{equation}
\alpha_2 = \frac{ \zeta ( 2 - d_w ) \Gamma ( 2 - d _w )}{\displaystyle
\zeta (1 + d_w) (2 \pi)^{2-d_w}} \cos \left ( \frac{\pi d_w}{2} \right ).
\end{equation}
We then compute the following large $r$ expansion of the scaled MFPT:
\begin{equation}
\tau(r) =  \frac{1}{\pi} \int_{0}^{\pi} \frac{1 - \cos \left ( k r \right )}{
\alpha_1 k^{d_w}} dk +  \mathcal{O}( r^{2 d_w-3}).
\end{equation}
For any value of $d_w \in \  ]1,2[$, we obtain the following expression of the
leading term:
\begin{eqnarray}
\tau(r) & \simeq & \frac{ r^{d_w-1}}{\alpha_1 \pi}  \int_{0}^{\infty} \frac{1
- \cos \left ( u \right )}{u^{d_w}} dk \nonumber \\
& \simeq & \frac{d_w  r^{d_w-1}}{\displaystyle 2 \alpha_1 \Gamma(1+d_w) \cos
\left ( \frac{\pi (2 - d_w)}{2} \right )}
\label{eq:Tapprox1}.
\end{eqnarray}
We therefore obtain exactly the result (\ref{scale}) derived in the continuous
limit, with the multiplicative constant given by (\ref{eq:A}). In particular
the leading term of the MFPT does not depend on the exact form of the discrete
time propagator, but only on its continuous limit. 

To assess the subleading term, we have to distinguish between the cases $d_w>
3/2$ and $d_w \le 3/2$. For the latter case, we obtain a correction term to
the scaled MFPT behaving as $\mathcal{O}(1)$,
\begin{equation}
\tau(r)= \frac{d_w}{2 \alpha_1 \Gamma(1+d_w) \cos \left ( \frac{\pi (2 -
d_w)}{2} \right )}  r^{d_w-1} + B +o(1),
\label{eq:Tapprox2}
\end{equation}
where the subleading term $B$, defining the residual MFPT, is a constant, that
does not vanish in the limit $L\to\infty$,
\begin{equation}\label{residualB1}
B  = \frac{1}{\pi}  \int_0^{\pi} \frac{1}{1- \lambda \left (k \right )}-
\frac{1}{\alpha_1 k^{d_w}}  dk -  \frac{1}{\alpha_1 (d_w-1) \pi^{d_w}}.
\end{equation}
The residual MFPT $B$ depends on the full $k$ expansion of $\lambda (k)$, and
thus on the precise form of the propagator. If the leading term is the same
for all jump processes sharing the same limiting propagator, the residual
term depends on the shape of the jump distribution, as was found in the case
of jump processes with finite second moment. The validity of expansion
(\ref{eq:Tapprox2}) is verified numerically in Fig. \ref{fig:LevyT.75}.

\begin{figure}[htb]
\includegraphics[width =0.5\linewidth,clip]{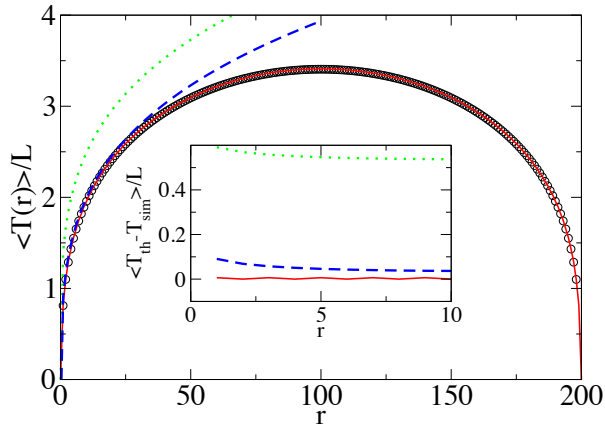} 
\caption{Mean First Passage Time for a L{\'e}vy flight on a $1D$ lattice of
size 200, with $d_w = 1.25$, in discrete time (black circles) compared to the
theoretical expression of Eq.~(\ref{eq:Texact}) (red line), to the rough
approximation (\ref{eq:Tapprox1}) (green dotted line) and to the
approximation (\ref{eq:Tapprox2}) (blue dashed line). The inset shows the
difference between the three theoretical expressions ($T_{\rm th}$) and the simulation ($T_{\rm sim}$).}
\label{fig:LevyT.75}
\end{figure}

\begin{figure}[htb]
\includegraphics[width =0.5\linewidth,clip]{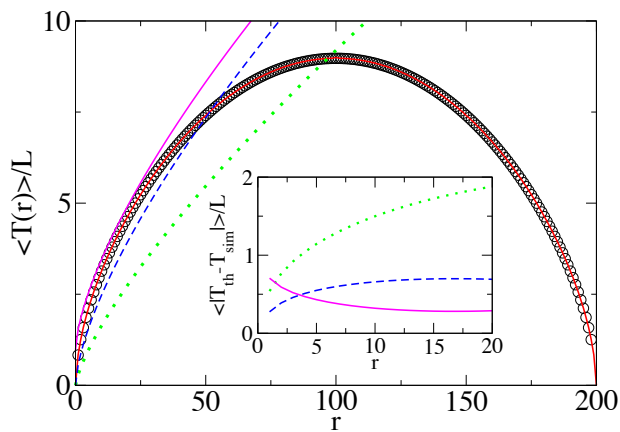} 
\caption{Mean First Passage Time for a L{\'e}vy flight on a network of size 200,
for $d_w = 1.75$, in discrete time (black circles) compared to the theoretical
expression (\ref{eq:Texact}) (red line), to the rough approximation of
(\ref{eq:Tapprox1}) (green dotted line), and to the approximation
(\ref{eq:Tapprox3}) with the first subleading term (blue dashed line) and
with two subleading terms (magenta plain line). The inset shows the
difference between the three theoretical expressions ($T_{\rm th}$) and the simulation ($T_{\rm sim}$).}
\label{fig:LevyT.752}
\end{figure}

For the case $d_w>3/2$ the third term in the $k$ expansion of $\lambda (k)$
has to be taken into account. We now need to compute a second subleading term
in the large $r$ expansion,
\begin{widetext}
\begin{eqnarray}
\tau(r) & = & \frac{1}{\pi} \int_{0}^{\pi} \frac{1 - \cos \left ( k r \right
)}{\alpha_1 k^{d_w}} dk + \frac{1}{\pi} \int_{0}^{\pi}  \frac{1 - \cos \left
( k r \right )}{\displaystyle \frac{\alpha_1^2}{\alpha_2} k^{2 d_w - 2}} dk
+ B+o(1)\\
& = & \frac{d_w}{2 \alpha_1 \pi \Gamma(1+d_w) \cos \left ( \frac{\pi(2-d_w)}{2}
\right )}  r^{d_w-1} + \frac{(2 d_w -2) \alpha_2}{2 \alpha_1^2 \Gamma(2d_w-1)
\cos ( \pi d_w )}  r^{2 d_w-3} +  B+ o(1).
\label{eq:Tapprox3}
\end{eqnarray}
Importantly, this expansion yields again a residual term $B$, which does not
vanish for small $r$, and which depends on the full jump distribution:
\begin{equation}
\label{residualB}
B=\frac{1}{\pi}\int_0^{\pi} \left(\frac{1}{1- \lambda \left (k \right )}-
\frac{1}{\alpha_1 k^{d_w}}-\frac{\alpha_2}{\alpha_1^2 k^{2 d_w - 2}}\right)dk
-\frac{1}{\alpha_1 (d_w-1) \pi^{d_w}}-\frac{\alpha_2}{\alpha_1^2(2d_w-3)\pi^{
2 d_w-2}}.
\end{equation}
\end{widetext}
The validity of expansion (\ref{eq:Tapprox3}) is verified numerically in
Fig.~\ref{fig:LevyT.752}.

To emphasize how critically the residual MFPT depends on the jump distribution,
we consider the example of a L{\'e}vy flight in discrete time, but now in
continuous space. At each time step, the random walker makes a step distributed
according to a symmetric L{\'e}vy stable distribution of index $\alpha=d_w$.
The structure function $\lambda (k)$ is then directly the characteristic
function of the L{\'e}vy distribution,
\begin{equation}\label{lambdalevy}
\lambda (k) = \exp (- | ck|^{d_w})=1-(ck)^{d_w}-\frac{(ck)^{2 d_w}}{2}+
\mathcal{O}(k^{2 d_w}).
\end{equation}
The leading term of the MFPT is then the continuous limit obtained in
Eqs.~(\ref{asympcontinuous}) and (\ref{Alevy}), with $\alpha_1 = c^{d_w}$.
However, the subleading term in this case is always a constant residual MFPT,
whose value depends on the specific shape of the jump distribution. This means
that the scaled MFPT admits different large $r$ expansions for Riemann walks
(distribution (\ref{eq:distrib})) and continuous space L{\'e}vy flights, which
depend on the full jump distribution and not only its asymptotics. Importantly,
we found that in both cases there exists a non-vanishing residual correction to
the MFPT. These results, together with previous section, suggest that a
non-vanishing residual MFPT exists as soon as the jump distribution allows
leapovers.

\section{Splitting probabilities}

\begin{figure}[htb!]
\includegraphics[width = 0.5\linewidth,clip]{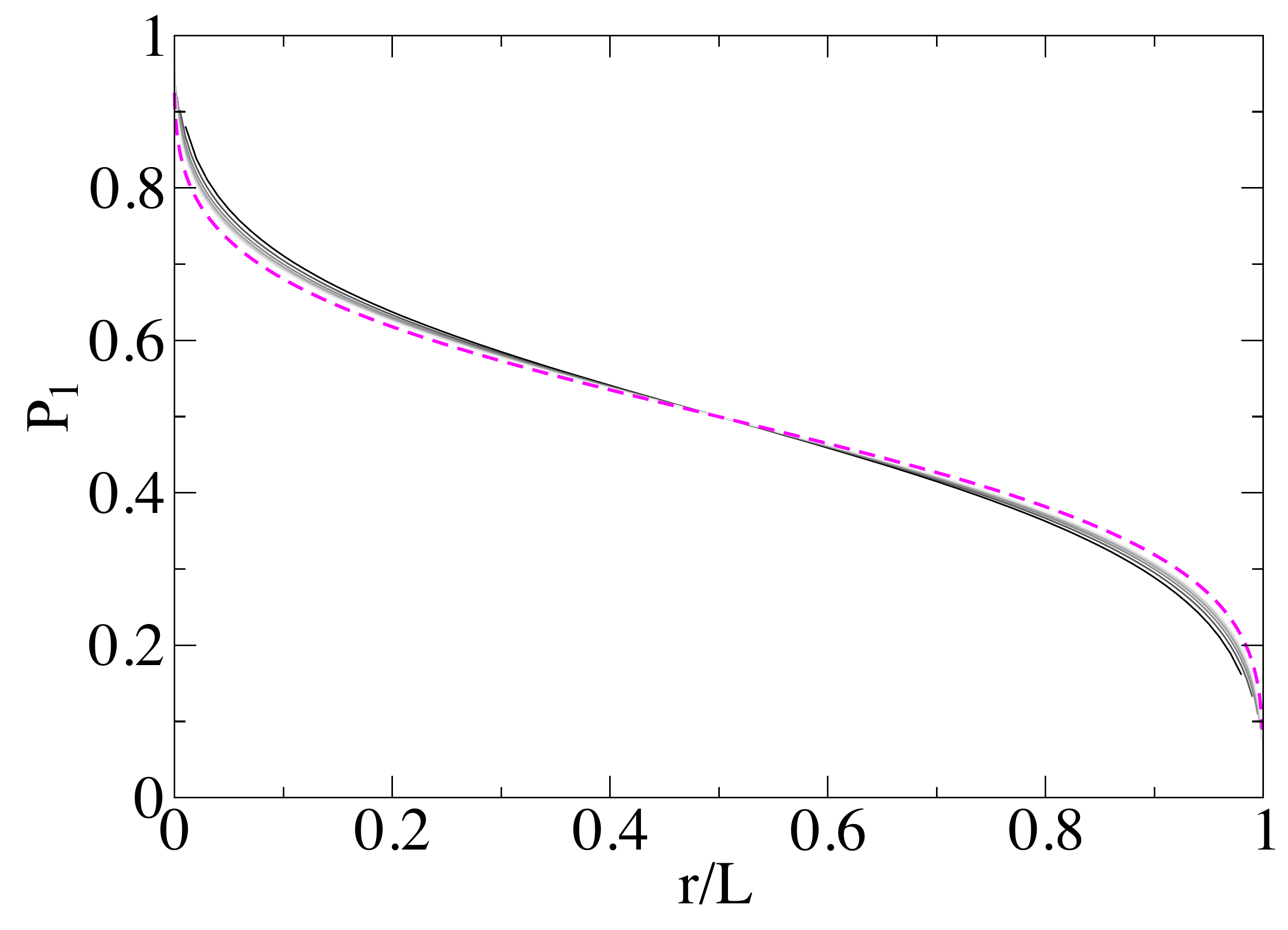} 
\caption{Splitting probability for L{\'e}vy flights, for several sizes of the
discrete network (200, 400, 800, 1600, 3200 and 6400 sites, from black to light grey as the size grows) compared to the theoretical
expression of Eq. (\ref{eq:P1}) (magenta dotted line), and for $d_w=1.25$.}
\label{fig:LevyP.25}
\end{figure}

\begin{figure}[htb!]
\includegraphics[width = 0.5\linewidth,clip]{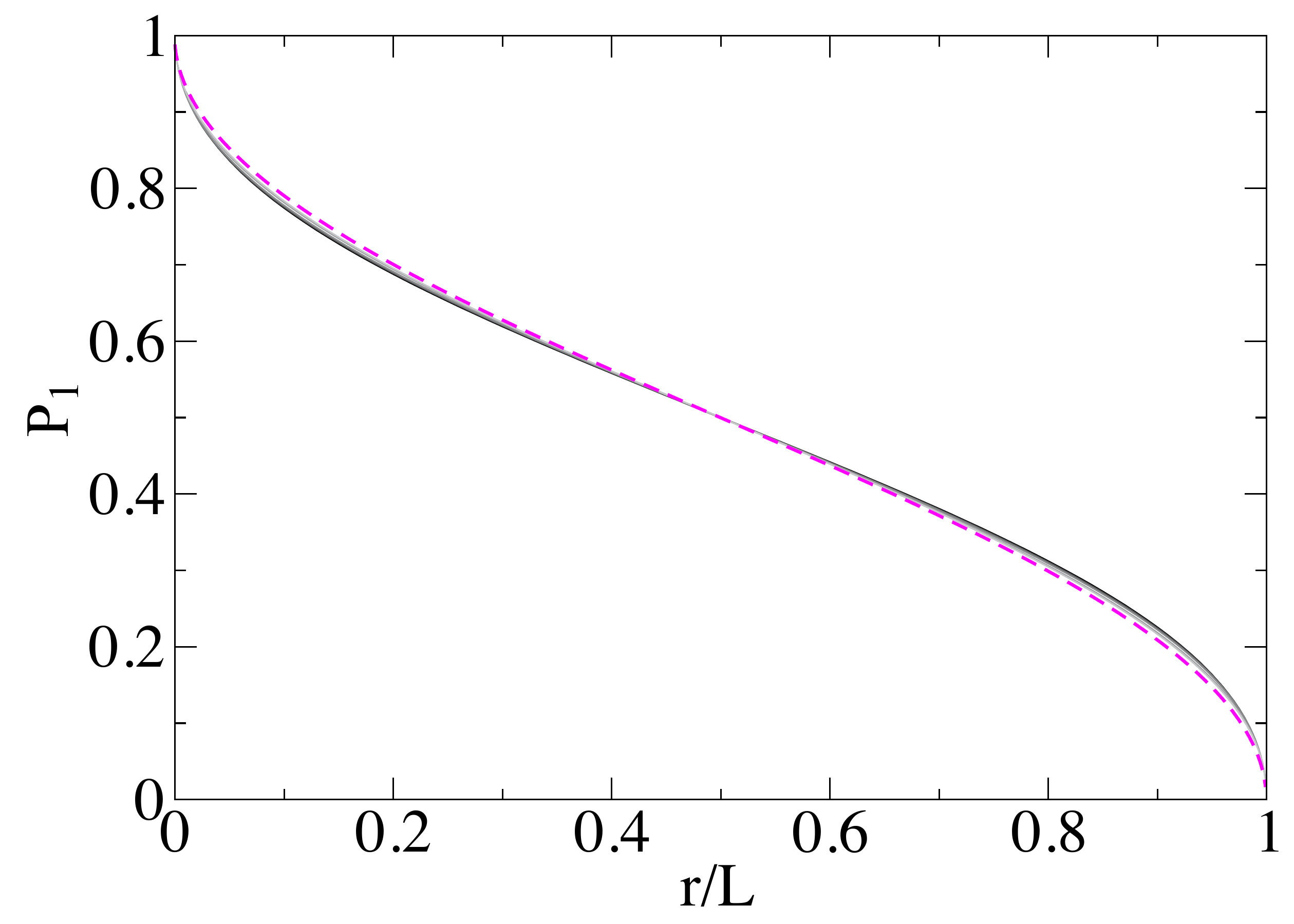} 
\caption{Splitting probability for L{\'e}vy flights, for several size of the
discrete network (200, 400, 800, 1600, 3200 and 6400 sites, from black to light grey as the size grows) compared to the theoretical
expression of Eq. (\ref{eq:P1}) (magenta dotted line), and for $d_w=1.5$.}
\label{fig:LevyP.5}
\end{figure}

\begin{figure}[htb!]
\includegraphics[width = 0.5\linewidth,clip]{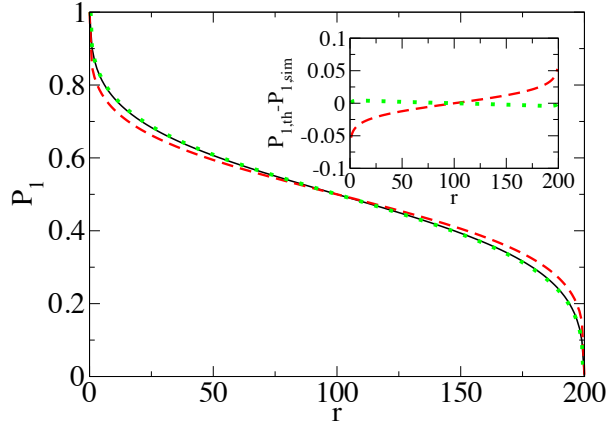} 
\caption{Splitting probability for L{\'e}vy flights on a network of size 200,
for $d_w=1.25$, in discrete time (black line), compared to the theoretical
expression (\ref{eq:P1}) (red dashed line), and to the exact expression (\ref{split}) where the first subleading term of the MFPT of Eq.~(\ref{residualB1})  is taken into account (green dotted line). 
The inset shows the difference between the two theoretical expressions ($P_{\rm1, th}$) and
the simulation ($P_{\rm1, sim}$).}
\label{fig:LevyP.75}
\end{figure}

The above formalism can be extended to further first-passage observables,
including observables involving several targets, following the method
developed in Ref.~\cite{al:2011fk}. As an illustrative example we consider
the case of splitting probabilities. More precisely, we assume that the jump
process takes place on a ring of length $2L$, with a target $T_1$ at position
$r=0$ and a target $T_2$ at position $r=L$. Note that in the case of processes
with leapovers, first-passage problems involving $N>2$ targets do not reduce
to problems with two targets. We denote by $P_1$ the probability that the walker
hits $T_1$ before ever reaching $T_2$. It is known that $P_1$ can be exactly
expressed in terms of MFPTs as follows \cite{Condamin:2007eu},
\begin{equation}\label{split}
P_1 (r,L) = \frac{\langle {\bf T}(r+L,2L) \rangle - \langle {\bf T}(r,2L)
\rangle +\langle {\bf T}(L,2L) \rangle }{2 \langle {\bf T}(L,2L) \rangle}.
\end{equation}
Indeed, it can be easily seen that this form satisfies the boundary conditions
$P_1(0,L)=1$ and $P_1(L,L)=0$, as well as the backward equation $\Delta_r P_1(
r,L)=0$. The exact form of the MFPT derived in Eq.~(\ref{eq:Texact}) therefore
provides straightforwardly an exact expression of the splitting probabilities.
In particular, it admits a simple closed form in the continuous limit, which
reads
\begin{widetext}
\begin{equation}
P_1\left(x=\frac{r}{L}\right)=\frac{1}{2}\frac{\zeta(1-d_w,\frac{1+x}{2})+
\zeta(1-d_w,\frac{1-x}{2})-\zeta (1 - d_w,\frac{x}{2})-\zeta (1- d_w,1-
\frac{x}{2})}{\displaystyle  2 \zeta (1 - d_w,\frac{1}{2})-\zeta (1 - d_w,0)
-\zeta (1-d_w,1) } + \frac{1}{2}.
\label{eq:P1} 
\end{equation}
\end{widetext}
For $d_w=2$ (Brownian case), we retrieve the classical expression, using
$\zeta (-n,x) = -B_{n+1}(x)/(n+1)$, where $B_n$ are the Bernouilli polynomials
($B_2(x) = x^2-x+1/6$),
\begin{eqnarray*}
P_1 (x) & = & \frac{B_2\left ( \frac{x}{2} \right ) +B_2\left (1- \frac{x}{2}
\right ) -B_2\left ( \frac{1+x}{2} \right ) -B_2\left ( \frac{1-x}{2} \right)}{
2 B_2(0) +2 B_2(1) - 4 B_2\left ( \frac{1}{2} \right )}+\frac{1}{2}\\
& = & 1-x.
\end{eqnarray*}
In the generic case $d_w\not=2$, Eq.~(\ref{eq:P1}) provides an exact expression
for the splitting probability of a $1D$ L{\'e}vy process. We compare this exact
solution in the continuous limit (\ref{eq:P1}) with numerical simulations of
L{\'e}vy flights. The results are shown in Figs.~\ref{fig:LevyP.25},
\ref{fig:LevyP.5}, and \ref{fig:LevyP.75}, indicating a rather slow convergence
to the continuous limit as the system size grows. Clearly, the residual MFPT
analyzed in previous sections yields a non vanishing correction to the splitting
probabilities in the large system size limit. Fig.~\ref{fig:LevyP.75} shows that
taking into account the residual MFPT in the evaluation of the MFPTs entering
Eq.~(\ref{split}), significantly improves the results for finite size $L$ in
the regimes of small $r$ and small $L-r$. These results can be straightforwardly
generalized to a larger number of targets using the formalism developed in
Ref.~\cite{al:2011fk}.

\section{MFPT for Fractional Brownian Motion}

\begin{figure}
\includegraphics[width =0.5\linewidth,clip]{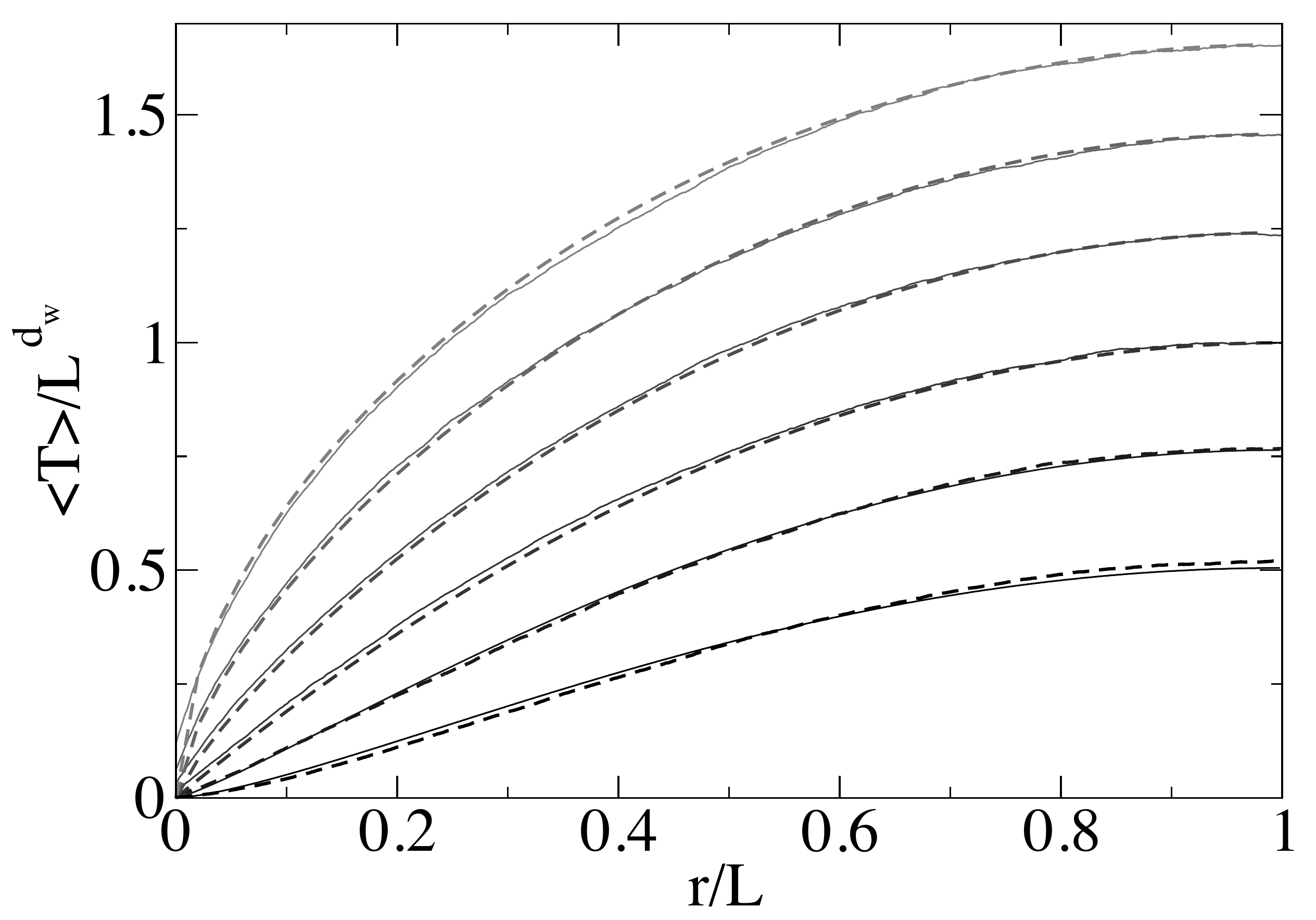} 
\caption{MFPT for an FBM, for several values of the Hurst exponent $H = 1/d_w$
($0.40$, $0.45$, $0.50$, $0.55$, $0.60$ and $0.65$, from black [bottom] to light grey [top] as $H$ grows). The continuous lines represents the simulations
results, and the dashed lines the theoretical expressions following
Eq.~(\ref{eq:f4}), where $A$ is a fitting parameter.}
\label{fig:FBMT}
\end{figure}

Expressions (\ref{eq:f4}) and (\ref{eq:P1}) are valid for any continuous,
compact, self-similar and Markovian $1D$ random walk. Fractional Brownian
motion (FBM) shares all these properties except for the Markov condition.
FBM is defined as a continuous-time gaussian process with zero mean and
stationary increments. More specifically, FBM is defined through the
Langevin equation
\begin{equation}
\frac{dx(t)}{dt}=\xi(t)
\end{equation}
for the position $x(t)$, which is driven by the stationary, fractional
Gaussian noise $\xi(t)$ with $\langle\xi(t)\rangle=0$ and long-ranged
noise correlation \cite{mandelbrot:422}
\begin{equation}
\label{fgn}
\langle\xi(t_1)\xi(t_2)\rangle=2H(2H-1)|t_1-t_2|^{2H-2}+4H|t_1-t_2|^{2H-1}
\delta(t_1-t_2),
\end{equation}
where we chose a unity diffusion constant. This process has the following
covariance,
\begin{equation}
\langle x(t)x(s)\rangle=\frac{1}{2}\left(|t|^{2H}+|s|^{2H}-|t-s|^{2H}\right).
\end{equation}
For subdiffusive processes ($H<1/2$), the underlying fractional Gaussian noise
(\ref{fgn}) is anticorrelated, while it is positively correlated in the
superdiffusive case $1/2<H<1)$.

To test numerically the applicability of our above results to non-Markovian
processes, we performed simulations of FBM and compared the MFPT to a target
for this process with expression (\ref{eq:f4}). The results shown in
Fig.~\ref{fig:FBMT} demonstrate that expression (\ref{eq:f4}) with free
parameter $A$ provides a surprisingly good approximation of the MFPT for FBM,
even though this process is highly non-Markovian (see also the discussion in
Ref.~\cite{Sliusarenko:2010}). These numerical findings suggest that memory effects do
not play an overly crucial role in the determination of MFPTs for FBM, and that the
range of applications of the approach developed herein in practise extends to
examples of non-Markovian processes.

\section{Conclusion and summary}

To conclude, we obtained a functional equation for the MFPT of a generic
self-similar Markovian, continuous process to a target in a $1D$ domain,
and derived its exact solution. We showed that such a continuous limit of the
MFPT is actually different from the large system size limit for discrete jump
processes allowing leapovers. In the leapover case, the large system size limit
of the MFPT involves non-vanishing corrections, which we call residual MFPT.
This residual time can have important consequences in the context of both search
processes and  numerical simulations of first-passage times of random walks.
We have investigated in detail the case of L{\'e}vy flights, and validated our
results by numerical simulations. We also demonstrated numerically, that our
results apply with surprising accuracy to FBM, despite its non-Markovian
nature.

\acknowledgments

Financial support from the CompInt graduate school at TUM,  the Academy
of Finland (FiDiPro scheme) and Mines ParisTech are gratefully acknowledged.

\bibliographystyle{vincent}


\begin{thebibliography}{24}
\expandafter\ifx\csname natexlab\endcsname\relax\def\natexlab#1{#1}\fi
\expandafter\ifx\csname bibnamefont\endcsname\relax
  \def\bibnamefont#1{#1}\fi
\expandafter\ifx\csname bibfnamefont\endcsname\relax
  \def\bibfnamefont#1{#1}\fi
\expandafter\ifx\csname citenamefont\endcsname\relax
  \def\citenamefont#1{#1}\fi
\expandafter\ifx\csname url\endcsname\relax
  \def\url#1{\texttt{#1}}\fi
\expandafter\ifx\csname urlprefix\endcsname\relax\def\urlprefix{URL }\fi
\providecommand{\bibinfo}[2]{#2}
\providecommand{\eprint}[2][]{\url{#2}}




\bibitem[{\citenamefont{Hughes}(1995{\natexlab{b}})}]{Hughes1995}
\bibinfo{author}{\bibfnamefont{B.}~\bibnamefont{Hughes}},
  \emph{\bibinfo{title}{Random Walks and Random Environments}}
  (\bibinfo{publisher}{Oxford University Press, New York},
  \bibinfo{year}{1995}{\natexlab{b}}).




\bibitem[{\citenamefont{Redner}(2001)}]{Redner:2001}
\bibinfo{author}{\bibfnamefont{S.}~\bibnamefont{Redner}},
  \emph{\bibinfo{title}{A guide to First- Passage Processes}}
  (\bibinfo{publisher}{Cambridge University Press}, \bibinfo{year}{2001}).

\bibitem[{\citenamefont{Rice}(1985)}]{Rice:1985}
\bibinfo{author}{\bibfnamefont{S.~A.} \bibnamefont{Rice}},
  \emph{\bibinfo{title}{Diffusion-limited reactions}},
  vol.~\bibinfo{volume}{25} (\bibinfo{publisher}{Elsevier, Amsterdam},
  \bibinfo{year}{1985}).

\bibitem[{\citenamefont{H{\"a}nggi et~al.}(1990)\citenamefont{H{\"a}nggi,
  Talkner, and Borkovec}}]{Hanggi:1990a}
\bibinfo{author}{\bibfnamefont{P.}~\bibnamefont{H{\"a}nggi}},
  \bibinfo{author}{\bibfnamefont{P.}~\bibnamefont{Talkner}}, \bibnamefont{and}
  \bibinfo{author}{\bibfnamefont{M.}~\bibnamefont{Borkovec}},
  \bibinfo{journal}{Reviews of Modern Physics} \textbf{\bibinfo{volume}{62}}
  (\bibinfo{year}{1990}).

\bibitem[{\citenamefont{B\'enichou et~al.}(2008)\citenamefont{B\'enichou,
  Loverdo, Moreau, and Voituriez}}]{PCCP}
\bibinfo{author}{\bibfnamefont{O.}~\bibnamefont{B\'enichou}},
  \bibinfo{author}{\bibfnamefont{C.}~\bibnamefont{Loverdo}},
  \bibinfo{author}{\bibfnamefont{M.}~\bibnamefont{Moreau}}, \bibnamefont{and}
  \bibinfo{author}{\bibfnamefont{R.}~\bibnamefont{Voituriez}},
  \bibinfo{journal}{Physical Chemistry Chemical Physics}
  \textbf{\bibinfo{volume}{10}}, \bibinfo{pages}{7059} (\bibinfo{year}{2008}).

\bibitem[{\citenamefont{Lloyd and May}(2001)}]{Lloyd:2001}
\bibinfo{author}{\bibfnamefont{A.~L.} \bibnamefont{Lloyd}} \bibnamefont{and}
  \bibinfo{author}{\bibfnamefont{R.~M.} \bibnamefont{May}},
  \bibinfo{journal}{Science} \textbf{\bibinfo{volume}{292}},
  \bibinfo{pages}{1316} (\bibinfo{year}{2001}).

\bibitem[{\citenamefont{Benichou et~al.}(2005)\citenamefont{Benichou, Coppey,
  Moreau, Suet, and Voituriez}}]{Benichou:2005qd}
\bibinfo{author}{\bibfnamefont{O.}~\bibnamefont{Benichou}},
  \bibinfo{author}{\bibfnamefont{M.}~\bibnamefont{Coppey}},
  \bibinfo{author}{\bibfnamefont{M.}~\bibnamefont{Moreau}},
  \bibinfo{author}{\bibfnamefont{P.-H.} \bibnamefont{Suet}}, \bibnamefont{and}
  \bibinfo{author}{\bibfnamefont{R.}~\bibnamefont{Voituriez}},
  \bibinfo{journal}{Phys Rev Lett} \textbf{\bibinfo{volume}{94}},
  \bibinfo{pages}{198101} (\bibinfo{year}{2005}).

\bibitem[{\citenamefont{Benichou et~al.}(2006)\citenamefont{Benichou, Loverdo,
  Moreau, and Voituriez}}]{Benichou:2006lq}
\bibinfo{author}{\bibfnamefont{O.}~\bibnamefont{Benichou}},
  \bibinfo{author}{\bibfnamefont{C.}~\bibnamefont{Loverdo}},
  \bibinfo{author}{\bibfnamefont{M.}~\bibnamefont{Moreau}}, \bibnamefont{and}
  \bibinfo{author}{\bibfnamefont{R.}~\bibnamefont{Voituriez}},
  \bibinfo{journal}{Phys Rev E Stat Nonlin Soft Matter Phys}
  \textbf{\bibinfo{volume}{74}}, \bibinfo{pages}{020102}
  (\bibinfo{year}{2006}).

\bibitem[{\citenamefont{Shlesinger}(2006)}]{Shlesinger:2006}
\bibinfo{author}{\bibfnamefont{M.~F.} \bibnamefont{Shlesinger}},
  \bibinfo{journal}{Nature} \textbf{\bibinfo{volume}{443}},
  \bibinfo{pages}{281} (\bibinfo{year}{2006}).

\bibitem[{\citenamefont{Lomholt et~al.}(2008)\citenamefont{Lomholt, Tal,
  Metzler, and Joseph}}]{Lomholt:2008}
\bibinfo{author}{\bibfnamefont{M.~A.} \bibnamefont{Lomholt}},
  \bibinfo{author}{\bibfnamefont{K.}~\bibnamefont{Tal}},
  \bibinfo{author}{\bibfnamefont{R.}~\bibnamefont{Metzler}}, \bibnamefont{and}
  \bibinfo{author}{\bibfnamefont{K.}~\bibnamefont{Joseph}},
  \bibinfo{journal}{Proceedings of the National Academy of Sciences}
  \textbf{\bibinfo{volume}{105}}, \bibinfo{pages}{11055}
  (\bibinfo{year}{2008}).


  
  \bibitem[{\citenamefont{Condamin
  et~al.}(2007{\natexlab{b}})\citenamefont{Condamin, Benichou, Tejedor,
  Voituriez, and Klafter}}]{Condamin:2007zl}
\bibinfo{author}{\bibfnamefont{S.}~\bibnamefont{Condamin et al.}},
  \bibinfo{journal}{Nature} \textbf{\bibinfo{volume}{450}}, \bibinfo{pages}{77}
  (\bibinfo{year}{2007}{\natexlab{b}});
\bibinfo{author}{\bibfnamefont{O.}~\bibnamefont{Benichou}},
  \bibinfo{author}{\bibfnamefont{B.}~\bibnamefont{Meyer}},
  \bibinfo{author}{\bibfnamefont{V.}~\bibnamefont{Tejedor}}, \bibnamefont{and}
  \bibinfo{author}{\bibfnamefont{R.}~\bibnamefont{Voituriez}},
  \bibinfo{journal}{Phys Rev Lett} \textbf{\bibinfo{volume}{101}},
  \bibinfo{pages}{130601} (\bibinfo{year}{2008});
\bibinfo{author}{\bibfnamefont{O.}~\bibnamefont{Benichou}} \bibnamefont{and}
  \bibinfo{author}{\bibfnamefont{R.}~\bibnamefont{Voituriez}},
  \bibinfo{journal}{Phys Rev Lett} \textbf{\bibinfo{volume}{100}},
  \bibinfo{pages}{168105} (\bibinfo{year}{2008}).
  
  

\bibitem[{\citenamefont{Tejedor et~al.}(2009)\citenamefont{Tejedor, B\'enichou,
  and Voituriez}}]{Tejedor:2009vn}
\bibinfo{author}{\bibfnamefont{V.}~\bibnamefont{Tejedor}},
  \bibinfo{author}{\bibfnamefont{O.}~\bibnamefont{B\'enichou}},
  \bibnamefont{and}
  \bibinfo{author}{\bibfnamefont{R.}~\bibnamefont{Voituriez}},
  \bibinfo{journal}{Physical Review E} \textbf{\bibinfo{volume}{80}}
  (\bibinfo{year}{2009}).

\bibitem[{\citenamefont{B{\'e}nichou et~al.}(2010)\citenamefont{B{\'e}nichou,
  Chevalier, Klafter, Meyer, and Voituriez}}]{BenichouO.:2010a}
\bibinfo{author}{\bibfnamefont{O.}~\bibnamefont{B{\'e}nichou}},
  \bibinfo{author}{\bibfnamefont{C.}~\bibnamefont{Chevalier}},
  \bibinfo{author}{\bibfnamefont{J.}~\bibnamefont{Klafter}},
  \bibinfo{author}{\bibfnamefont{B.}~\bibnamefont{Meyer}}, \bibnamefont{and}
  \bibinfo{author}{\bibfnamefont{R.}~\bibnamefont{Voituriez}},
  \bibinfo{journal}{Nat Chem} \textbf{\bibinfo{volume}{2}},
  \bibinfo{pages}{472} (\bibinfo{year}{2010}).
  
  
  
  
  \bibitem{havlin} D.Ben-Avraham and S.Havlin, {\it  Diffusion and reactions in fractals and disordered systems} (Cambridge University Press, 2000).
 
 
 \bibitem{bouchaud} J.-P. Bouchaud and A. Georges, Phys. Rep. \textbf{195}, 127
(1990).

\bibitem{mekla} R. Metzler and J. Klafter, Phys. Rep. \textbf{339}, 1 (2000).


 
\bibitem[{\citenamefont{R.Metzler and J.Klafter}(2004)}]{R.Metzler:2004}
\bibinfo{author}{\bibnamefont{R.Metzler}} \bibnamefont{and}
  \bibinfo{author}{\bibnamefont{J.Klafter}}, \bibinfo{journal}{J.Phys.A}
  \textbf{\bibinfo{volume}{37}}, \bibinfo{pages}{R161} (\bibinfo{year}{2004}).

\bibitem[{\citenamefont{Chechkin et~al.}(2003)\citenamefont{Chechkin, Metzler,
  Gonchar, Klafter, and Tanatarov}}]{Chechkin:2003}
\bibinfo{author}{\bibfnamefont{A.~V.} \bibnamefont{Chechkin}},
  \bibinfo{author}{\bibfnamefont{R.}~\bibnamefont{Metzler}},
  \bibinfo{author}{\bibfnamefont{V.~Y.} \bibnamefont{Gonchar}},
  \bibinfo{author}{\bibfnamefont{J.}~\bibnamefont{Klafter}}, \bibnamefont{and}
  \bibinfo{author}{\bibfnamefont{L.~V.} \bibnamefont{Tanatarov}},
  \bibinfo{journal}{Journal of Physics A: Mathematical and General}
  \textbf{\bibinfo{volume}{36}}, \bibinfo{pages}{L537} (\bibinfo{year}{2003}).

\bibitem[{\citenamefont{Koren et~al.}(2007)\citenamefont{Koren, Lomholt,
  Chechkin, Klafter, and Metzler}}]{Metzler:2007}
\bibinfo{author}{\bibfnamefont{T.}~\bibnamefont{Koren}},
  \bibinfo{author}{\bibfnamefont{M.~A.} \bibnamefont{Lomholt}},
  \bibinfo{author}{\bibfnamefont{A.~V.} \bibnamefont{Chechkin}},
  \bibinfo{author}{\bibfnamefont{J.}~\bibnamefont{Klafter}}, \bibnamefont{and}
  \bibinfo{author}{\bibfnamefont{R.}~\bibnamefont{Metzler}},
  \bibinfo{journal}{Phys. Rev. Lett.} \textbf{\bibinfo{volume}{99}},
  \bibinfo{pages}{160602} (\bibinfo{year}{2007}).

\bibitem[{\citenamefont{P{\'o}lya}(1920)}]{Polya:1920}
\bibinfo{author}{\bibfnamefont{G.}~\bibnamefont{P{\'o}lya}},
  \bibinfo{journal}{Mathematische Zeitschrift} \textbf{\bibinfo{volume}{8}},
  \bibinfo{pages}{171} (\bibinfo{year}{1920}),
  \bibinfo{note}{10.1007/BF01206525}.

\bibitem[{\citenamefont{Gnedenko and Kolmogorov}(1954)}]{Kolmogorov:1954}
\bibinfo{author}{\bibfnamefont{B.~V.} \bibnamefont{Gnedenko}} \bibnamefont{and}
  \bibinfo{author}{\bibfnamefont{A.~N.} \bibnamefont{Kolmogorov}},
  \emph{\bibinfo{title}{Limit Distributions for Sums of Independent Random
  Variables}} (\bibinfo{publisher}{Addison-Wesley}, \bibinfo{year}{1954}).



\bibitem[{\citenamefont{Montroll and Weiss}(1965)}]{Montroll:1965}
\bibinfo{author}{\bibfnamefont{E.~W.} \bibnamefont{Montroll}} \bibnamefont{and}
  \bibinfo{author}{\bibfnamefont{G.~H.} \bibnamefont{Weiss}},
  \bibinfo{journal}{Journal of Mathematical Physics}
  \textbf{\bibinfo{volume}{6}}, \bibinfo{pages}{167} (\bibinfo{year}{1965}).



\bibitem[{\citenamefont{Chevalier et~al.}(2011)\citenamefont{Chevalier,
  Benichou, Meyer, and Voituriez}}]{al:2011fk}
\bibinfo{author}{\bibfnamefont{C.}~\bibnamefont{Chevalier}},
  \bibinfo{author}{\bibfnamefont{O.}~\bibnamefont{Benichou}},
  \bibinfo{author}{\bibfnamefont{B.}~\bibnamefont{Meyer}}, \bibnamefont{and}
  \bibinfo{author}{\bibfnamefont{R.}~\bibnamefont{Voituriez}},
  \bibinfo{journal}{Journal of Physics A: Mathematical and Theoretical}
  \textbf{\bibinfo{volume}{44}} (\bibinfo{year}{2011}).

\bibitem[{\citenamefont{Condamin
  et~al.}(2007{\natexlab{b}})\citenamefont{Condamin, Benichou, and
  Moreau}}]{Condamin:2007eu}
\bibinfo{author}{\bibfnamefont{S.}~\bibnamefont{Condamin}},
  \bibinfo{author}{\bibfnamefont{O.}~\bibnamefont{Benichou}}, \bibnamefont{and}
  \bibinfo{author}{\bibfnamefont{M.}~\bibnamefont{Moreau}},
  \bibinfo{journal}{Phys Rev E Stat Nonlin Soft Matter Phys}
  \textbf{\bibinfo{volume}{75}}, \bibinfo{pages}{021111}
  (\bibinfo{year}{2007}{\natexlab{b}}).

\bibitem[{\citenamefont{Mandelbrot and Ness}(1968)}]{mandelbrot:422}
\bibinfo{author}{\bibfnamefont{B.~B.} \bibnamefont{Mandelbrot}}
  \bibnamefont{and} \bibinfo{author}{\bibfnamefont{J.~W.~V.}
  \bibnamefont{Ness}}, \bibinfo{journal}{SIAM Review}
  \textbf{\bibinfo{volume}{10}}, \bibinfo{pages}{422} (\bibinfo{year}{1968}).

\bibitem[{\citenamefont{Sliusarenko et~al.}(2010)\citenamefont{Sliusarenko,
  Gonchar, Chechkin, Sokolov, and Metzler}}]{Sliusarenko:2010}
\bibinfo{author}{\bibfnamefont{O.~Y.} \bibnamefont{Sliusarenko}},
  \bibinfo{author}{\bibfnamefont{V.~Y.} \bibnamefont{Gonchar}},
  \bibinfo{author}{\bibfnamefont{A.~V.} \bibnamefont{Chechkin}},
  \bibinfo{author}{\bibfnamefont{I.~M.} \bibnamefont{Sokolov}},
  \bibnamefont{and} \bibinfo{author}{\bibfnamefont{R.}~\bibnamefont{Metzler}},
  \bibinfo{journal}{Phys. Rev. E} \textbf{\bibinfo{volume}{81}},
  \bibinfo{pages}{041119} (\bibinfo{year}{2010}).

\end{thebibliography}

\end{document}